\documentclass[preprintnumbers,amsmath,amssymb,twocolumn]{revtex4}
\usepackage{graphicx}
\usepackage{dcolumn}
\usepackage{bm}
\usepackage{natbib}
\usepackage{physics}
\usepackage[caption=false]{subfig}

\begin{document}

\title{Coherent Microwave Control of a Nuclear Spin Ensemble at Room Temperature}

\author{P. Huillery$^{1}$}
\author{J. Leibold$^{1}$}
\author{T. Delord$^{1}$}
\author{L. Nicolas$^{1}$}
\author{J. Achard $^{2}$}
\author{A. Tallaire $^{2,3}$}
\author{G. H\'etet$^{1}$} 

\affiliation{$^1$Laboratoire de Physique de l'Ecole Normale Sup\'erieure, ENS, Universit\'e PSL, CNRS, Sorbonne Universit\'e, Universit\'e de Paris, Paris, France. 
$^{2}$ LSPM-CNRS, UPR 3407, Université Sorbonne Paris Nord, 99, avenue JB Clément, 93430 Villetaneuse, France
$^{3}$ IRCP,Ecole Nationale Supérieure de Chimie de Paris, 11, rue Pierre et Marie Curie, 75005 Paris, France
}

\begin{abstract}
We use nominally forbidden electron-nuclear spin transitions in nitrogen-vacancy (NV) centers in diamond to demonstrate coherent manipulation of a nuclear spin ensemble using microwave fields at room temperature.
We show that employing an off-axis magnetic field with a modest amplitude ($\approx 0.01$ T) at an angle with respect to the NV natural quantization axes is enough to tilt the direction of the electronic spins, and enable efficient spin exchange with the nitrogen nuclei of the NV center. We could then demonstrate fast Rabi oscillations on electron-nuclear spin exchanging transitions, coherent population trapping and polarization of nuclear spin ensembles in the microwave regime. 
Coupling many electronic spins of NV centers to their intrinsic nuclei offers full scalability with respect to the number of controllable spins and provides prospects for transduction.
In particular, the technique could be applied to long-lived storage of microwave photons and to the coupling of nuclear spins to mechanical oscillators in the resolved sideband regime.
\end{abstract}

\maketitle

Nuclear spins in solids form extremely well isolated quantum systems that have emerged as promising candidates for solid-state quantum information processing and quantum sensing \cite{Zhong2015,Muhonen2014,PhysRevLett.113.246801}. However, due to their high degree of isolation, nuclear spins initialization, control and read-out remain challenging. A successful strategy is to use their coupling to electron spins.
It was realized 70 years ago that nuclear polarization can be enhanced by taking advantage of the magnetic polarizability of nearby electron spins \cite{PhysRev.92.411,PhysRev.92.212.2,Solid1,Solid2,HENSTRA1988389} with direct impact in Nuclear Magnetic Resonance and Imaging (NMR and MRI) \cite{BECERRA199528,Gerfen1995,Hall930}. 

More advanced quantum control uses optically polarizable electron spins \cite{Gruber,Koehl2011} found in some materials. Notably, in the last two decades, there has been considerable success in the implementation of the electronic spin of the negatively charged Nitrogen-Vacancy (NV) centers in diamond interacting with surrounding nuclear spins.
Coherent control of single $^{13}$C nuclear spins using nearby NV centers has for instance enabled a wealth of quantum effects to be observed such as the realization of a two-qubits conditional quantum gate \cite{PhysRevLett.93.130501}, arbitrary quantum state transfer between electron and nuclear spins \cite{Dutt1312} or coherent population trapping of single nuclear spins \cite{PhysRevLett.116.043603}. Single NVs have also been coupled to few \cite{Childress281,Neumann1326,Jiang267,PhysRevLett.109.137601,PhysRevLett.109.137602, Zhao2012, Yun_2019}, or a bath of surrounding $^{13}$C nuclear spins \cite{Hanson352,PhysRevB.80.041201,Togan2011,vanderSar2012,PhysRevLett.113.137601}. Recently, quantum imaging of a 27-spin ensemble \cite{Abobeih2019} and a 10-qubits quantum register \cite{PhysRevX.9.031045} have been achieved. 

In contrast with the randomness of the location of most nuclear spins with respect to the NV centers, $^{14}$N spins are deterministically present on every center. This should in principle allow a uniform scaling of some of the aforementioned interactions to ensembles. In particular, it would be beneficial for quantum memories, enabling a $\sqrt{N}$ scaling of the single photon coupling efficiency \cite{PhysRevLett.84.5094}.
However, the transverse (flip-flop) coupling between the $^{14}$N nuclear and NV electron spins is weak. Strong DC magnetic fields, precisely tuned to Level Anti-Crossings (LAC) \cite{PhysRevLett.102.057403,PhysRevA.80.050302,PhysRevB.87.125207,PhysRevA.100.011801,PhysRevA.94.021401}, have enabled nuclear spin polarization \cite{PhysRevA.80.050302,PhysRevB.87.125207}, enhanced electron spin read-out fidelity \cite{PhysRevB.81.035205} and electron to nuclear spin quantum state transfer \cite{Fuchs2011}. Seeking fast and controllable coupling schemes, AC fields have been used to drive nominally forbidden transitions, hereafter called electron-nuclear spin transitions (ENST), which change both the electron and nuclear spins, using an off-axis magnetic field \cite{PhysRevB.79.075203}. ENST have been observed near the ground state LAC \cite{PhysRevB.47.8809,PhysRevB.47.8816,Wei_1999,PhysRevB.100.075204}, in the RF domain \cite{PhysRevB.92.020101,PhysRevResearch.2.023094} where they have shown to effectively increase the nuclear $g$-factor, and with optical fields at 4K \cite{PhysRevLett.124.153203}.

In this letter, we demonstrate coherent driving of ensembles of nuclear spins \textit{via} ENST in the microwave regime. We use it to realize coherent population trapping of a nuclear spin ensemble at room temperature and discuss applications in quantum information storage. We also present a new method for polarizing nuclear spins using off-axis magnetic fields and propose an application in the field of spin-mechanics.

\begin{figure*}[!ht]
  \centering \scalebox{0.14}{\includegraphics{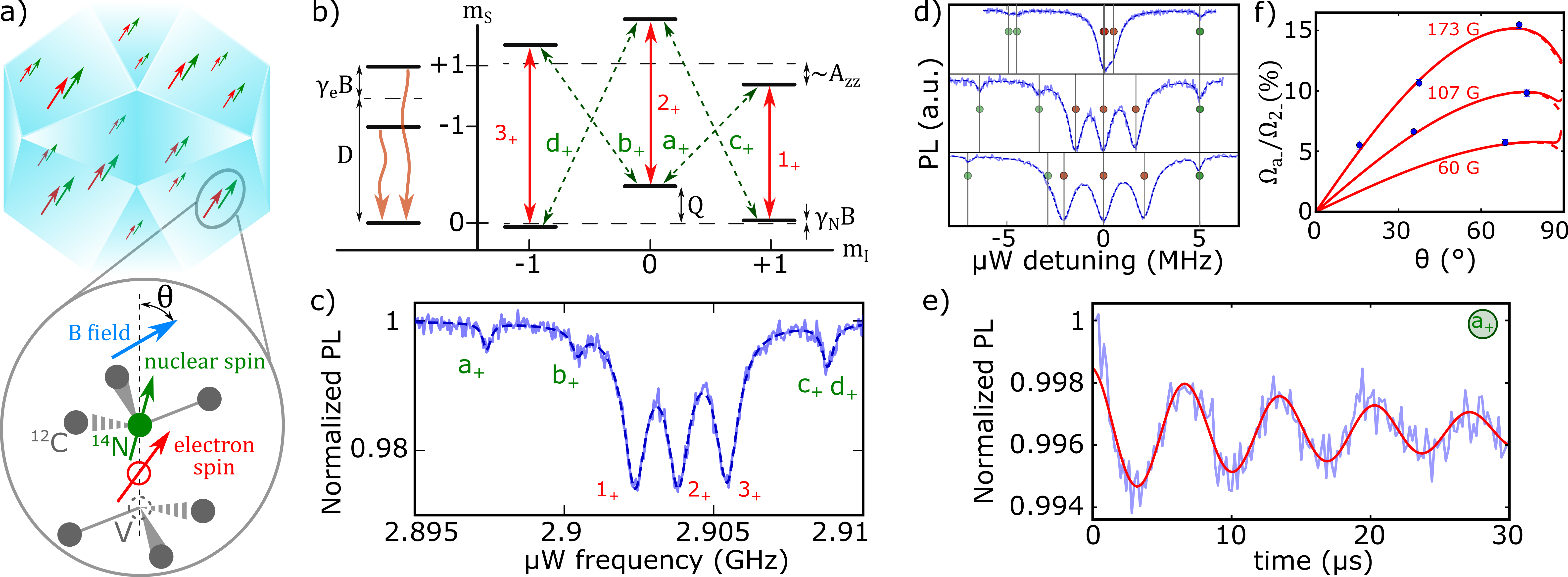}}
  \caption{a) Schematics showing a bulk diamond containing an ensemble of NV centers coupled to electronic and $^{14}$N nuclear spins. b) Energy level structure of a single NV center. Left : electron spin levels. Wavy arrows depict optical electron spin polarization to the $m_S=0$ state. Right : coupled electron-nuclear spins levels. Solid red arrows show nuclear spin preserving transitions ($1_+$, $2_+$ and $3_+$). Dashed green arrows show nuclear spin exchanging transitions ($a_+, b_+, c_+$ and $d_+$).  
 c) ODMR spectrum taken for $(B,\theta)=(75.0~{\rm G}, 87.8^{\circ}$), the dashed line is a guide to the eye. 
 d) ODMR spectra for $(B,\theta)=(75~{\rm G}, 89.7^{\circ}$) (top), $(B,\theta)=(75~{\rm G}, 87.8^{\circ}$) (middle) and $(B,\theta)=(57~{\rm G}, 84.0^{\circ}$), (bottom). Solid lines are experimental measurements, dashed lines are guides to the eye, points and vertical lines are theoretical predictions of the transition frequencies.
e) Rabi oscillations on transition $a_+$ for $(B,\theta)=(81~{\rm G}, 87.9^{\circ}$). The red solid line is an exponentially damped sinusoidal fit to the data. f) Ratio between the Rabi frequencies of the $a_-$ ($\Omega_{a_-}$) and  $2_-$transitions ($\Omega_{2_-}$) as a function of $\theta$ for different $B$-fields. Points are experimental data, solid lines (dashed lines) are theoretical predictions for a microwave field polarized along the $x$-axis ($y$-axis). 
 }
\end{figure*}

We consider an ensemble of spins in a bulk diamond as illustrated in Fig.1a. A single NV center can be described as composed of the $S=1$ electron spin and the $I=1$ $^{14}$N nuclear spin. Taking the NV direction as quantization axis, the NV center Hamiltonian reads 
\begin{eqnarray} \nonumber
H_{NV}/h &=& D\hat{S_z^2} + \gamma_e \hat{\textbf{S}}\cdot\textbf{B} + Q\hat{I_z^2} - \gamma_n \hat{\textbf{I}}\cdot\textbf{B} + \underline{\underline{\mathcal{A}}} \hat{\textbf{I}}\cdot\hat{\textbf{S}},
\end{eqnarray}
\noindent where $D=2.87$ GHz and $Q=-4.945$ MHz \cite{PhysRevA.80.050302} are the electron and nuclear spin zero field splittings , $\gamma_e = 2.802$ MHz/G and $\gamma_n = 0.308$ kHz/G are the electron and nuclear spin gyromagnetic factors respectively, and \underline{\underline{$\mathcal{A}$}} is the diagonal hyperfine interaction tensor with $\mathcal{A}_{zz} = -2.162$ MHz \cite{PhysRevA.80.050302} and $\mathcal{A}_{xx} = \mathcal{A}_{yy} = -2.62$ MHz \cite{PhysRevB.92.020101}. We neglect the effect of strain which is a good approximation for the high quality bulk diamond used in this work.

The level structure of the NV center is shown in Fig.1b, featuring the $3\times3$ quantum states of the two coupled spin $1$ systems labelled by the electron ($m_S = {-1,0,+1}$) and nuclear ($m_I = {-1,0,+1}$) magnetic quantum numbers. Under a B field along the NV axis, electron and nuclear spins states are not mixed : $m_S$ and $m_I$ are therefore good quantum numbers.
However, under an off-axis B field, $H_{NV}$ mixes the electron and nuclear spin states, leading to new eigenstates $\vert i \rangle = \sum_{m_S,m_I} \alpha_{m_S,m_I}^{i}\vert m_s,m_I \rangle$. As illustrated in Fig.1-a, the
 two spins indeed do not point to the same direction because of their differing zero field splitting ($D$ and $Q$) and gyromagnetic factors ($\gamma_e$ and $\gamma_n$). This authorizes ENST where both the electronic and nuclear spin states are changed, without significant reduction in the polarisation efficiency of the electronic spin for magnetic fields below one hundred Gauss (See SI).

Experimentally, we investigate this effect using a $^{12}$C enriched bulk diamond grown by Chemical Vapor Deposition (CVD). Injection of N$_2$ during the growing process gives a concentration of NV centers of  $\approx 0.3$~ppb in the diamond crystal. Using a confocal microscope (see SI), we perform Optically Detected Magnetic Resonance (ODMR) on an ensemble of hundreds of NV centers driven by a microwave signal.
Fig.1c shows an ODMR on one of the four NV classes at around 2.9 GHz under a magnetic field of $B=75$ G at an angle $\theta=87.8^{\circ}$ with respect to the NV axis (see SI for calibration details). Six electronic spin transitions from the $|m_s=0\rangle$ to the $|m_s=+1\rangle$ state (labelled with subscripts $+$) can clearly be observed, as described in Fig.1b. Similar transitions from the $|m_s=0\rangle$ to the $|m_s=-1\rangle$ state (labelled with subscripts $-$) are also observed (see SI). Those ENST in the microwave domain have not been reported in the literature although they have been shown to contribute to a modulation in electron spin echo signals \cite{PhysRevB.89.205202}.
Fig.1d shows ODMR spectra where $\theta$ is tuned towards $\pi/2$ (from bottom to top). Vertical lines on each graph are the $H_{NV}$ eigenfrequencies calculated numerically, showing very good agreement with the data.

Fig.1e is a measurement of the normalized NV photoluminescence as a function of microwave duration on the $a_+$ transition for $(B,\theta)=(81~{\rm G}, 87.9^{\circ}$), demonstrating coherent microwave driving of nuclear spins ensembles. 
We measure a Rabi frequency $\Omega_{a_+}=2\pi\times$147(1) kHz and a damping time $\tau_R=22(4)\mu s$ similar to the damping measured on the nuclear spin preserving transition (see SI). The Rabi frequency here is on the order of the RF Rabi frequencies measured close to NV level anti-crossings \cite{PhysRevA.80.050302,PhysRevB.87.125207}. 
The relative Rabi strengths $\Omega_{a_-}/\Omega_{2_-}$ for different B field amplitudes and orientations are plotted in Fig.1f. 
Theoretical calculations of the Rabi frequencies on the $\vert i \rangle \leftrightarrow \vert j \rangle$ transition, $\Omega_{ij} = \gamma_e B_{\mu W} \langle i \vert S_{x(y)} \vert j \rangle$ with $B_{\mu W}$ the B field amplitude at the spin position (see SI), show very good agreement with the data.
Note that for the B field magnitudes considered in the work, $\Omega_{n_p}/\Omega_{2_p}\approx 10~\%$ for $n={a,b,c,d}$ and $p={+,-}$. In order to avoid excitation of the nearby nuclear spin preserving transitions, an upper bound for $\Omega_{n_p}/2\pi$ is thus $\approx$ 100 kHz. 
\begin{figure}[h]
  \centering \scalebox{0.14}{\includegraphics{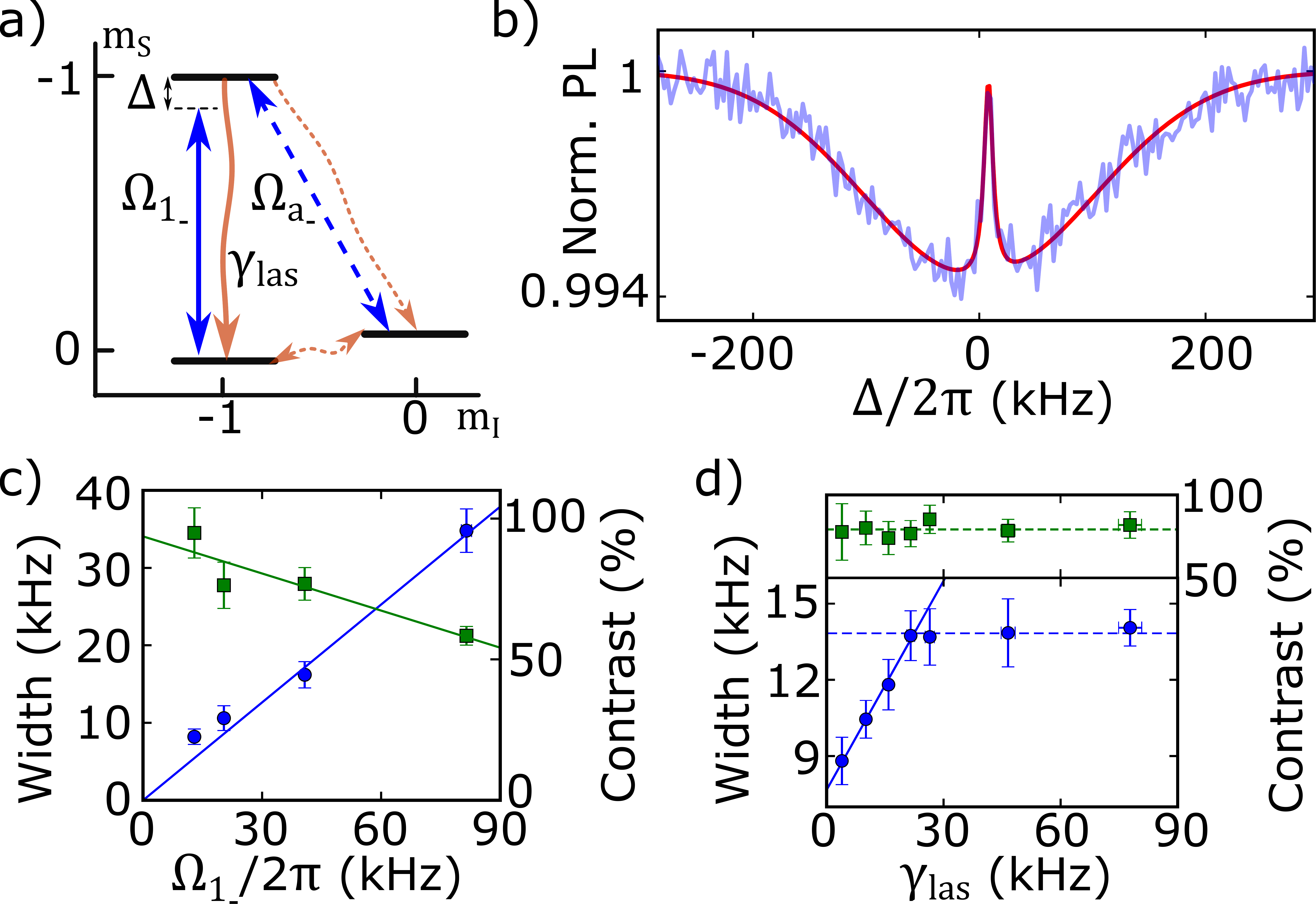}}
  \caption{a) Selected $\Lambda$-scheme in the coupled electron-nuclear spin level structure. Dashed wavy arrows represent the optically induced nuclear spin state relaxation. b) ODMR as a function of $\Delta/2\pi$ showing CPT with a nuclear spin ensemble. The blue/red line is the data/fit. c) Full Width at Half Maximum (FWHM) (blue circles, scale on the left) and contrast (green square, scale on the right) of the CPT peak are plotted as a function of $\Omega_{1_-}/2\pi$. Here, $\gamma_{las}$=18(1) kHz and $\Omega_{a_-}$=2$\pi\times$30.6(7) kHz. Solid lines are linear fits to the data. For the CPT width, the two first points are excluded from the fit. d) FWHM (blue circles, scale on the left) and contrast (green squares, scale on the right) of the CPT peak as a function of $\gamma_{las}$. Here, $\Omega_{1_-}$=2$\pi\times$24.7(2) kHz, $\Omega_{a_-}$=2$\pi\times$33.7(7) kHz. Solid lines are linear fits to the data, dashed lines are mean values of the data. $(B,\theta)=(82.71~{\rm G}, 10.2^{\circ}$) for all data.}
\end{figure}

\begin{figure*}
  \centering \scalebox{0.14}{\includegraphics{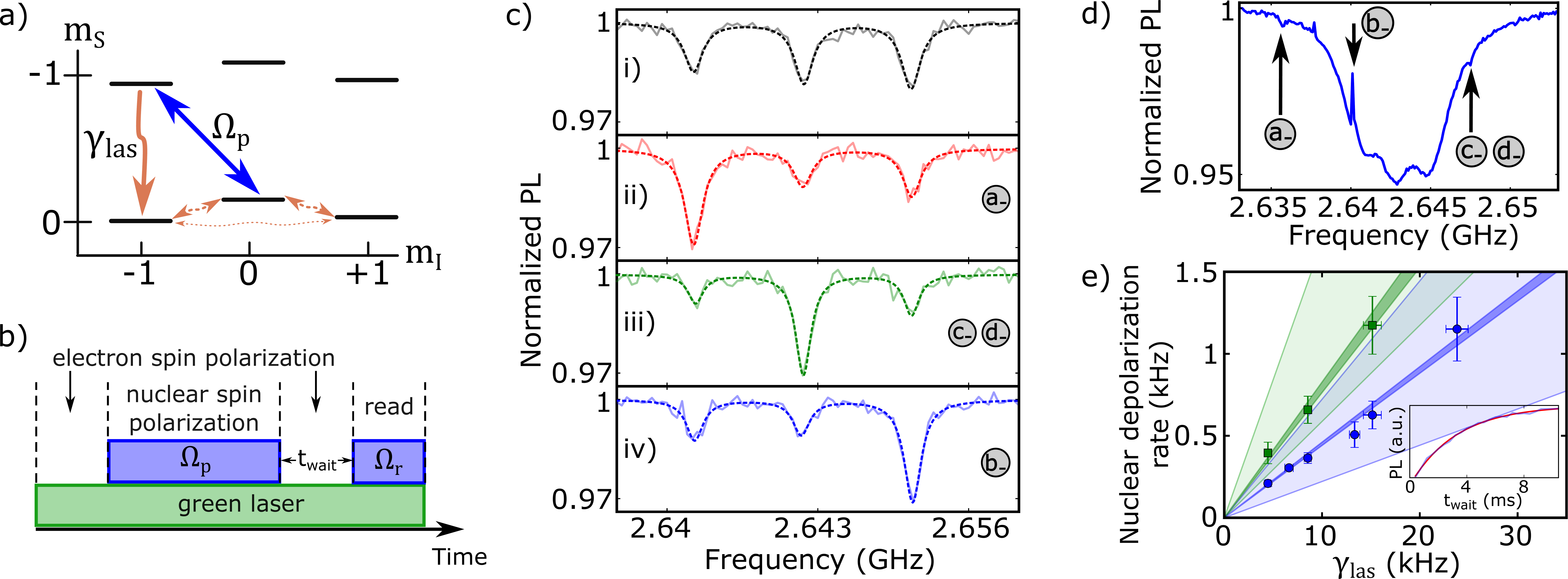}}
  \caption{a) Schematic of the nuclear spin polarization mechanism.  b) Preparation and measurement sequence. c) ODMR spectra without pump (i), with a resonant pump on the $a_-$, $c_-/d_-$ and $b_-$ transitions for traces ii), iii) and iv) respectively.
  $\gamma_{\rm las}=22.5(1.1)$~kHz, $\Omega_p$=2$\pi\times$17.0(4) kHz and the durations of the steps of the sequence are $(100,300,100$ and $25)\mu s$ (explanations in the text). Solid lines are data, dashed lines are guides to the eyes. d) ODMR spectrum taken at $(B,\theta)=(82.71~{\rm G}, 10.2^{\circ}$). Nuclear spin exchanging transitions are indicated by black arrows.  e) Depolarization rate of the nuclear spins from the $\vert m_S = 0, m_I = + 1 \rangle$ (blue circles) and $\vert m_S = 0, m_I = 0 \rangle$ (green squares) states versus electron spins polarization rate $\gamma_{las}$. Dots are data, solid lines are linear fit to the data. The inset is a typical nuclear spin depolarization measurement (blue line) with an exponential fit to the data (red line). The faint areas are results from a numerical model described in the SI.}
\end{figure*}

A direct consequence of these ENST is that three-level $\Lambda$-schemes can be isolated, offering the unprecedented possibility to realize Coherent Population Trapping (CPT) with nuclear spin ensembles at room temperature.
In a three-level $\Lambda$-scheme driven by two fields, the superposition of the two ground states, a so called dark state, is decoupled from the driving fields \cite{ARIMONDO1996257}. A consequence of this is that when the two fields are resonant, there is no population in the excited state in the steady-state across a narrow frequency window. Fig.2a depicts the $\Lambda$-scheme that we isolate experimentally. Fig.2b displays the measured ODMR spectrum.
Here, we set the laser repolarisation rate $\gamma_{\rm las}$ to $18(1)$ kHz (see SI), $\Omega_{a_-, 1_-}=2\pi\times (30.6(7), 12.9(2))$ kHz and $(B,\theta)=(82.71~{\rm G}, 10.2^{\circ}$).
As expected, a sharp peak appears close to two-photon resonance, with an almost full suppression of the population in the excited state at $\Delta/2\pi\approx 0$kHz.  Quantitatively, this characteristic CPT peak is well approximated by a Lorentzian with a width of 2$\pi\times$8.2(1.0) kHz and a contrast of 95(9)\% signaling very small nuclear spin relaxation. 

In atomic CPT, the transitions are in the optical domain, so the dissipative preparation of the dark state occurs through spontaneous emission. Interestingly, in the case of NV centers, dissipation is tunable {\it via} the green laser that polarizes the NV spin \cite{PhysRevLett.116.043603, Nicolas_2018}. One consequence of this is nuclear spin dephasing and relaxation induced by the optical excitation \cite{PhysRevLett.100.073001,Wang_2015, Neumann542,Blok2014}. 
Here, the electron-nuclear spin state mixing due to off-axis magnetic fields differs between ground and excited states, similar to when NV centers couple to the nuclear spin of proximal $^{13}$C atoms \cite{Dutt1312,PhysRevLett.110.060502,PhysRevLett.102.057403}.
Phenomenologically, two main resulting mechanisms are at play : decay from the $\vert m_S = -1, m_I = -1 \rangle$ state towards the $\vert m_S = 0, m_I = 0 \rangle$ state and population transfer between the $\vert m_S = 0, m_I = -1 \rangle$ and the $\vert m_S = 0, m_I = 0 \rangle$ states (see Fig.2a). 
Fig.2c is a measurement of the width of the CPT peak as a function of $\Omega_{1_-}/2\pi$, featuring a linear dependency of the width and contrasts as a function of Rabi frequency up to $90$kHz. This is at odds with standard CPT \cite{ARIMONDO1996257} where the widths and contrasts are expected to depend quadratically on the Rabi frequency.
In Fig.2d, we plot the same quantities as a function of the electron spin polarization rate $\gamma_{las}$. While the contrast remains constant at a value of $\approx 80~\%$, we observe that the CPT peak width first increases linearly with $\gamma_{las}$ 
and then reaches a plateau at a value of around 14 kHz for $\gamma_{las}$ above 22 kHz. This scaling also differs from the standard CPT scaling laws, where the width is expected to decrease with increasing relaxation \cite{ARIMONDO1996257}. 
Full modeling of the dynamics involving the 21 involved levels goes beyond the scope of this work, but these observations indicate that under strong optical illumination, dephasing and relaxation processes impact the CPT dynamics (see SI).

We now turn to another consequence of the ENST, namely the possibility to polarize ensembles of nuclear spins. As illustrated in Fig.3a, the electron spin polarization can in principle be transferred to the nuclear spins {\it via} a microwave pump $\Omega_p$. 
The employed experimental sequence is shown in Fig.3b. First, the electron spins are polarized in the $\vert m_S = 0 \rangle$ state. Then, the microwave pump is turned on. Finally, we let the electron spins re-polarize in the $\vert m_S = 0 \rangle$ state before applying a read-out microwave pulse with a smaller Rabi frequency $\Omega_r$. This sequence is repeated for different frequencies of the read-out pulse to measure the population in the different nuclear spin states.  Fig.3c ii-iii-iv) show three ODMR spectra obtained with the pump tuned to transitions $a_-$, $c_-/d_-$ and $b_-$ respectively and at an angle $\theta=10.2^\circ$, while trace i) is recorded with the pump off, demonstrating polarization of the whole $^{14}$N nuclear spin ensemble.

Unexpectedly, tuning the pump to transition $a_-$ resp. $b_-$ depopulates the nuclear spins out of the $\vert m_I=+1\rangle$ resp. $ \vert m_I= -1\rangle$ states, so that the whole process in the end feeds population back mostly in the desired nuclear spin state. This observation points towards a faster decay channel from the $\vert 0\rangle$ to the $\vert \pm1\rangle$ states than between the $\vert \pm 1\rangle$ states as indicated by the dashed red arrows in Fig.3a. 
We confirm this by measuring the nuclear spin depolarisation rate (inset of fig.3e) as a function of $\gamma_{las}$ from the state $\vert 0\rangle$ ($\gamma_{las}^{\vert 0\rangle}$, green squares in fig.3e) and $\vert +1\rangle$ ($\gamma_{las}^{\vert +1\rangle}$, blue circles) by recording the PL rate versus $t_{\rm wait}$. We first observe that $\gamma_{las}^{\vert +1\rangle}$ and $\gamma_{las}^{\vert 0\rangle}$ evolve linearly with $\gamma_{las}$ and thus with the optical power, as expected. Moreover, we find that $\gamma_{las}^{\vert 0\rangle}$/$\gamma_{las}$=8.0(3)\% is higher than $\gamma_{las}^{\vert +1\rangle}$/$\gamma_{las}$=4.5(1)\% giving a ratio $\gamma_{las}^{\vert 0\rangle}/\gamma_{las}^{\vert +1\rangle} = 1.8(1)$. As detailed in the SI, numerical modeling (faint areas in fig.3e) gives loosely bounded but compatible results, with $\gamma_{las}^{\vert 0\rangle}/\gamma_{las}^{\vert +1\rangle} = 2.27(3)$.
Overall, the presented results show a new method to selectively polarize all three nuclear spin states with a degree of polarization of $\approx 60\%$. Uniquely, this methods does not require a B field align with the NV axis. This new method could in principle be used with the $^{15}$N isotope with an even better polarisation since it only has two spin eigenstates.  

Let us finally discuss applications emerging from this work. 
In general, our observations open a path towards coupling ensembles of nuclear spins to degrees of freedom that are controllable by electronic spins. One example where such transduction could be employed is in spin-mechanics \cite{2017JOpt...19c3001L}, in particular using the libration of levitating diamonds or magnets \cite{Huillery,DelordNat} coupled identically to ensembles of NVs {\it via} off-axis B-fields. 
One outstanding difficulty of these experiments is to reach the so called resolved sideband regime, where the mechanical oscillator frequency $\Omega_m/2\pi$
exceeds the electronic spin resonance decay rate, a crucial step towards spin-cooling to the motional ground state. 
In this endeavor, nuclear spins can play a decisive role since they are isolated from their environment and feature very low decay rates. 
Nuclear spins have already been envisioned as a pristine system for spin-cooling cantilevers \cite{Greenberg, Cao, Okazaki}. However these proposals typically require low temperatures or LAC at large fields for nuclear spin-polarisation.
The presented ENST can be advantageously employed for resolved sideband manipulations using levitating diamonds at room temperature. The nuclear spins can indeed be polarized using the present scheme in the ideal $B$-field angle and magnitude for spin-mechanical coupling \cite{DelordNat}. A subsequent tone can then drive the many long-lived nuclear spins on the resolved red motional sideband at a frequency $Q-\Omega_m/2\pi$ without requiring a large mechanical frequency. Estimations suggest that using $10^{10}$ nuclear spins at pressures below $10^{-3}$ mbars will then cool the mechanical oscillator to the motional ground state.

A second promising direction would be light storage using Electromagnetically-Induced Transparency (EIT), the counterpart of CPT. In addition to ultra-narrow spectral features and its applications in metrology, EIT is the physical playground for the standard quantum memory protocol \cite{PhysRevLett.84.5094} which have been now realized using a large variety of systems. 
There are widespread studies on the strong coupling of electronics spins to microwaves cavities at ambient conditions \cite{Breeze}. 
For now, a quantum memory for microwaves using NV centers has been realized using photon-echo techniques on an electron spin transition \cite{PhysRevA.85.012333,PhysRevX.4.021049,PhysRevA.92.020301} at cryogenic conditions. Our work thus opens up clear perspectives for the use of EIT-based memories for storing microwave photons in a nuclear spin ensemble at room temperature.

In conclusion, we have shown that the $^{14}$N nuclear spin of NV centers can be coherently manipulated by microwave fields in the presence of an off-axis B field. We used this to realize CPT with a nuclear spin ensemble at room temperature and to demonstrate a new method to polarize nuclear spins under small magnetic fields. Our results will have important implications as transducers for protocols that require long-lived spin ensembles. 

\begin{acknowledgements}
We would like to acknowledge fruitful discussions with Y. Chassagneux, V. Jacques and A. Dr\'eau. This work has been supported by Region Ile-de-France in the framework of the DIM SIRTEQ. GH acknowledges funding by the T-ERC program through the project QUOVADIS.
\end{acknowledgements}

\onecolumngrid
\setcounter{figure}{0}
\renewcommand{\thefigure}{S\arabic{figure}}

\vspace{0.5in}
\begin{center}
{\Large \textsc{Supplementary Material} }
\end{center}

\subsection{Experimental setup}

\begin{figure}[!ht]
  \centering \scalebox{0.7}{\includegraphics{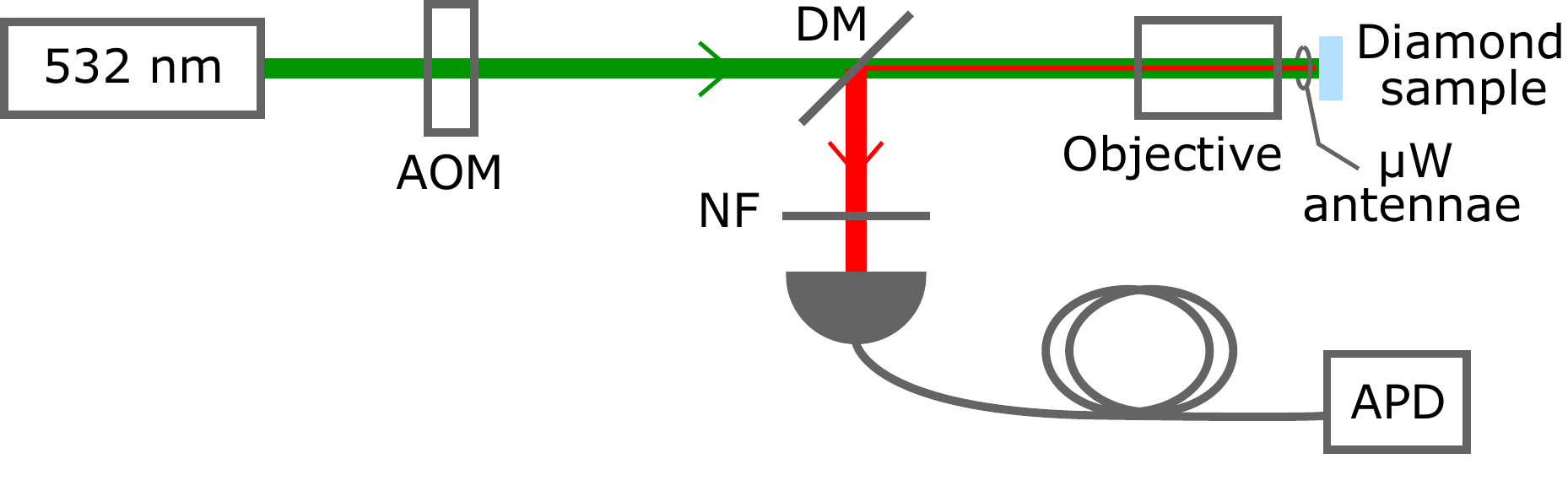}}
  \caption{Optical setup of the experiment.}
	\label{Optics}
\end{figure}

As illustrated in Fig.\ref{Optics}, the diamond sample is typicaly illuminated with 1mW of 532 nm laser light, focused by a NA = 0.75 objective (MPLN50x from Olympus). An acousto-optic modulator (AOM) is used to switch on and off the 532nm laser and to finely tuned its power. The photo-luminescence (PL) is collected by the objective, separated form the excitation light using a dichroic mirror (DM) and a 532nm notch filter (NF), and detected using a multimode-fibered single-photon avalanche photo-detector (APD) (SPCM-AQRH-15 from Perkin Elmer). Typically, we detect PL photons at a rate of 1MHz. 

A $\approx$ 1mm diameter single loop antenna is placed near the diamond sample (between the objective and the sample) to apply a microwave field to the NV centers. In the experiments where a single-tone microwave is needed, the signal is generated with a Rohde \& Schwarz SMB100A RF generator, sent to a switch (ZASWA-2-50-DR+  from  Mini-Circuits) and then to an amplifier (ZHL-15W-422-S+  from  Mini-Circuits) before feeding the antenna. When a two-tone microwave field is required, a second signal is generated by a SG4400L RF generator from DS Instruments, sent to a switch (ZASWA-2-50-DR+  from  Mini-Circuits) and combined with the first signal before the amplifier using a power combiner (ZAPD-4-S+ from Mini-Circuits). A computer-controlled card (PulseBlaster from SpinCore Technologies, Inc.) is used to generate the TTL pulses sent to the switches, to trigger the microwave frequency changes and the photon detection allowing the whole setup to be synchronized.

\subsection{Magnetic field calibration}

\begin{figure}[!ht] 
  \centering \scalebox{0.8}{\includegraphics{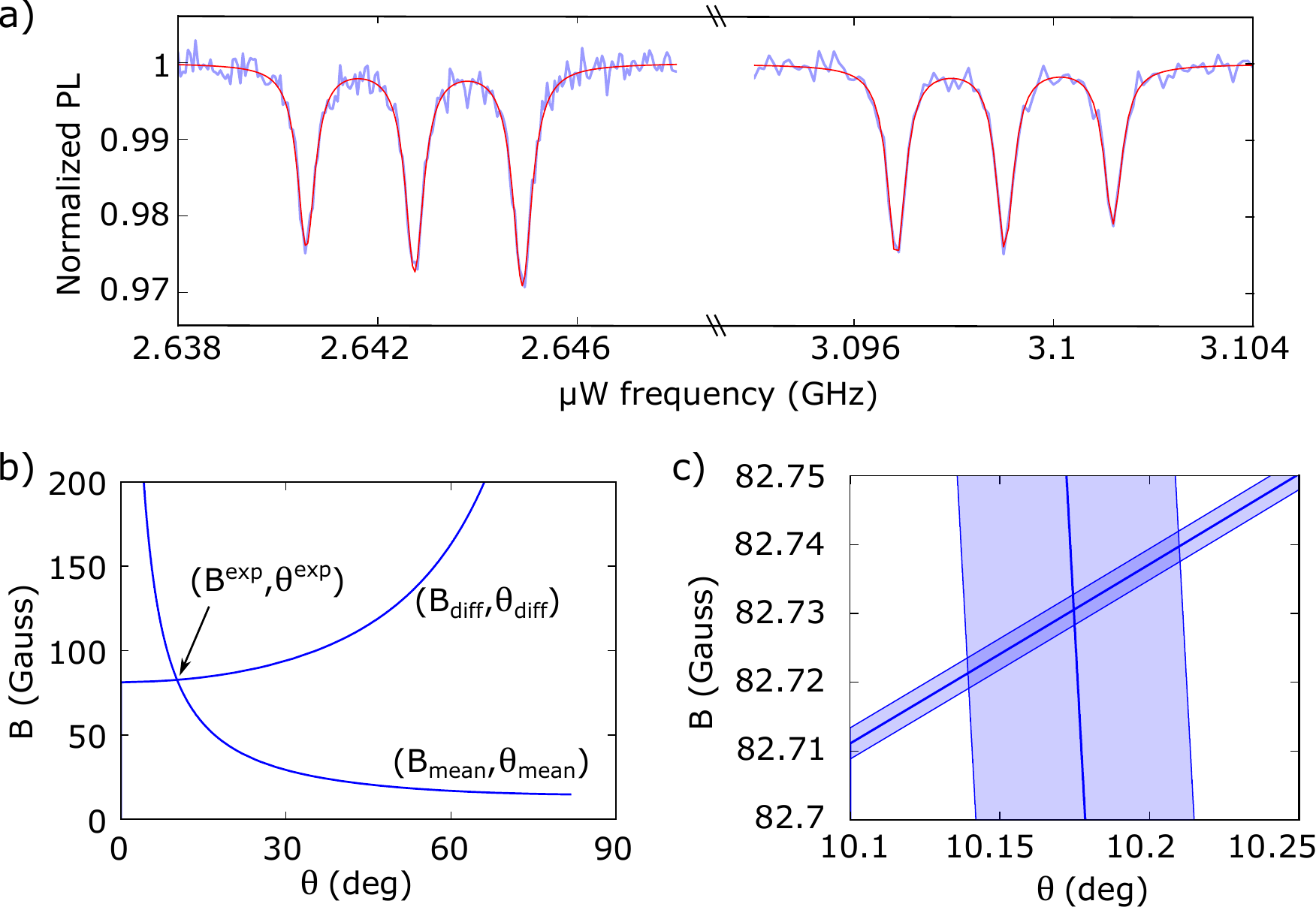}}
  \caption{a) ODMR spectra showing the $\vert m_S=0 \rangle \leftrightarrow \vert m_S=-1 \rangle$ and $\vert m_S=0 \rangle \leftrightarrow \vert m_S=+1 \rangle$ transitions. b) Graphical determination of $(B^{exp},\theta^{exp})$. c) Zoom on the crossing point with experimental uncertainties.}
	\label{FigCalib}
\end{figure}

A permanent magnet is placed a few cm away from the diamond sample in order to apply a uniform magnetic field to the NV centers.
To calibrate the magnetic field magnitude $B$, and its orientation $\theta$ with respect to the NV axis, we record Optically Detected Magnetic Resonace (ODMR) spectra. We measure both the $\vert m_S=0, m_I=0 \rangle \leftrightarrow \vert m_S=-1, m_I=0 \rangle$ and $\vert m_S=0, m_I=0 \rangle \leftrightarrow \vert m_S=+1, m_I=0 \rangle$ transition frequencies (see Fig.\ref{FigCalib}a). We then perform reverse engineering on the NV center electron spin Hamiltonian to deduce $B$ and $\theta$ from the transition frequencies measurement.

For a range of ($B$,$\theta$) values, we diagonalize numerically $H_{e}/h = D\hat{S_z^2} + \gamma_e \hat{\textbf{S}}\cdot\textbf{B}$ 
to calculate the frequency difference $E_{diff}(B,\theta)$ between and the mean frequency $E_{mean}(B,\theta)$ of the $\vert m_S=0 \rangle \leftrightarrow \vert m_S=+1 \rangle$ and $\vert m_S=0 \rangle \leftrightarrow \vert m_S=-1 \rangle$ transitions. For a given value $E_{diff(mean)}^{exp}$, one can determine all possible pairs $(B_{diff(mean)},\theta_{diff(mean)})$ such that $E_{diff(mean)}(B_{diff(mean)},\theta_{diff(mean)}) = E_{diff(mean)}^{exp}$. As shown in Fig.\ref{FigCalib}b), $(B_{diff},\theta_{diff})$ and $(B_{mean},\theta_{mean})$ intercept at one point. One can thus graphically determine the values $(B^{exp},\theta^{exp})$ corresponding to a given pair of experimentally measured ($E_{diff}^{exp}$,$E_{mean}^{exp}$).

As shown in Fig.\ref{FigCalib}c), the experimental uncertainties of the transition frequency measurements translate to uncertainties for the determination of $B$ and $\theta$. Precisions of 0.1 Gauss and 0.1 degree are typically obtained. For values quoted in the main text, uncertainties are on the order of one unit of the last significant digit.

\subsection{Rabi oscillations on the nuclear spin preserving and direct transitions}

\begin{figure}[!ht] 
  \centering \scalebox{0.5}{\includegraphics{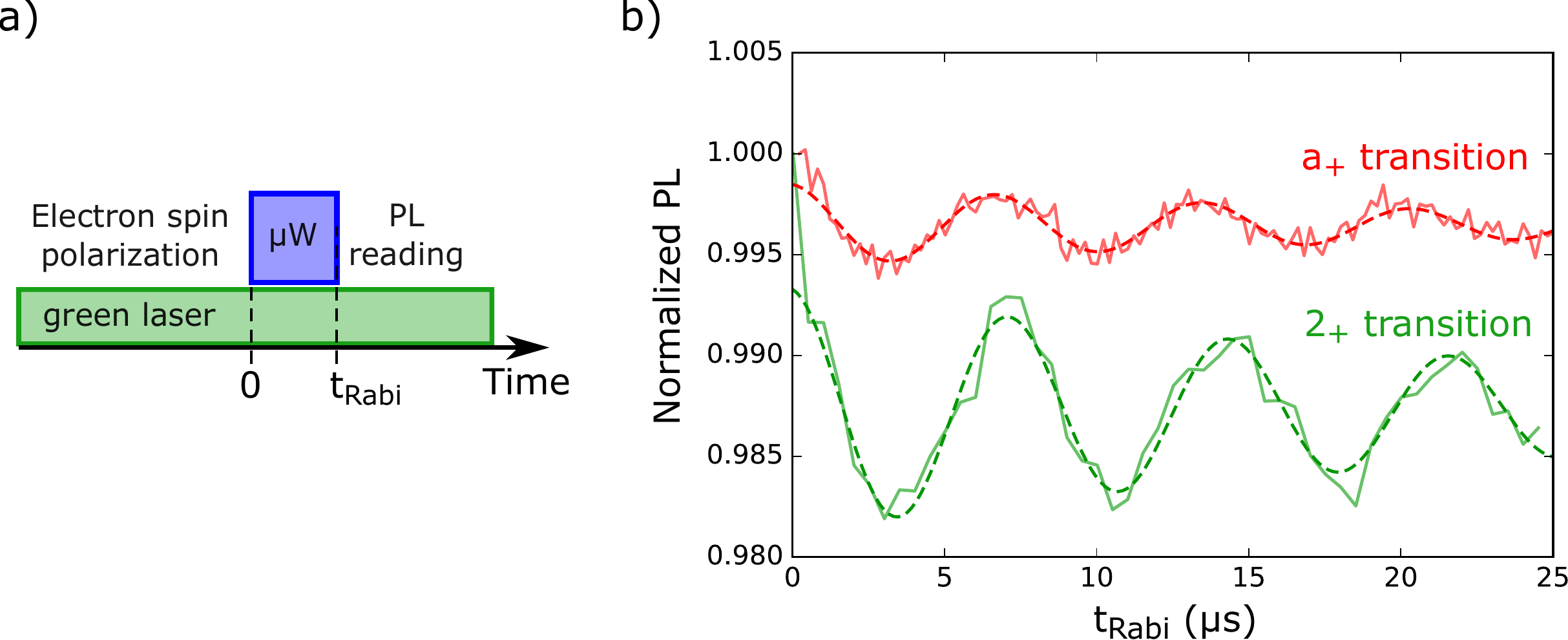}}
  \caption{a) Sequence used for the Rabi oscillations measurements. b) Rabi oscillations for the $a-$transition (in red) and the 2-transition (in green). Solid lines are the data, dashed lines are exponentially damped sinusoidal fits to the data.}
	\label{Fig-rabi}
\end{figure}

In this section, we describe the experimental measurements of Rabi oscillations shown in Fig. 1 in the main text. The same sequence, shown in Fig.\ref{Fig-rabi}a), is used for both nuclear spin exchanging and preserving transitions. In Fig.\ref{Fig-rabi}b), we plot Rabi oscillations for both the $a-$transition and the 2-transition. 

In these measurements, the Rabi frequency for the 2-transition ($\Omega_2=2\pi\times$138(1) kHz) is similar than for the $a-$transition ($\Omega_a=2\pi\times$147(1) kHz). Damping times $\tau_{R}=26(5)\mu$s and 22(4) $\mu$s are measured, for the 2 and a transitions respectively, indicating that there is no appreciable difference between the dephasing of the nuclear spin exchanging and preserving transitions.

\subsection{Measurement of the electron spin $T_2^{*}$ time}

\begin{figure}[!ht] 
  \centering \scalebox{0.45}{\includegraphics{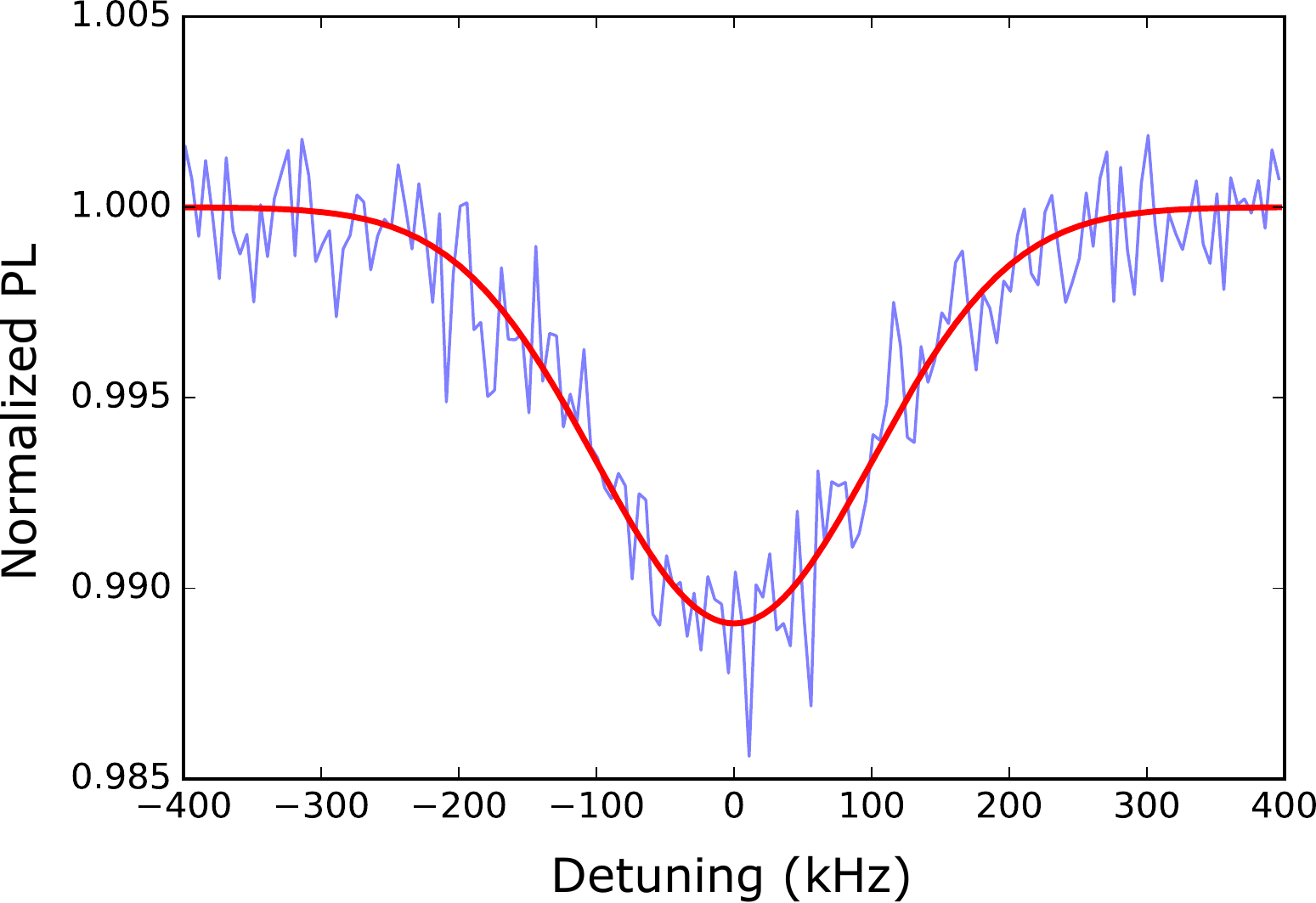}}
  \caption{ODMR spectrum of a single electron spin resonance taken at low driving field. The data (blue) are fitted by a Gaussian function (red).}
	\label{ESR-line}
\end{figure}

The Rabi decay time $\tau_{R}$ can be compared to the electronic spin dephasing time $T_2^{*}$ (given mostly by the interaction between NV and the $P_1$ centers in the sample). The latter can be extracted from the width of an electron spin resonance at low driving field. We show such an ODMR spectrum in Fig.\ref{ESR-line}. Fitting the data by a Gaussian lineshape, we measure a full width at half maximum $\Gamma_2^* =$ 237(8) kHz. Using the relation $\Gamma_2^* = \frac{2\sqrt{ln2}}{\pi T_2^*}$ \cite{SIdreau} gives $T_2^{*} = $ 2.2(1) $\mu$s. 
Note that since $\Omega_a \tau_R> 1$ here, continuous dynamical decoupling enhances $\tau_R$ above $T_2^*$.

\subsection{Measurement of the nuclear spin $T_1$ time in the presence of the green laser}

\subsubsection{NV center electron spin polarisation}

\begin{figure}[!ht] 
  \centering \scalebox{0.6}{\includegraphics{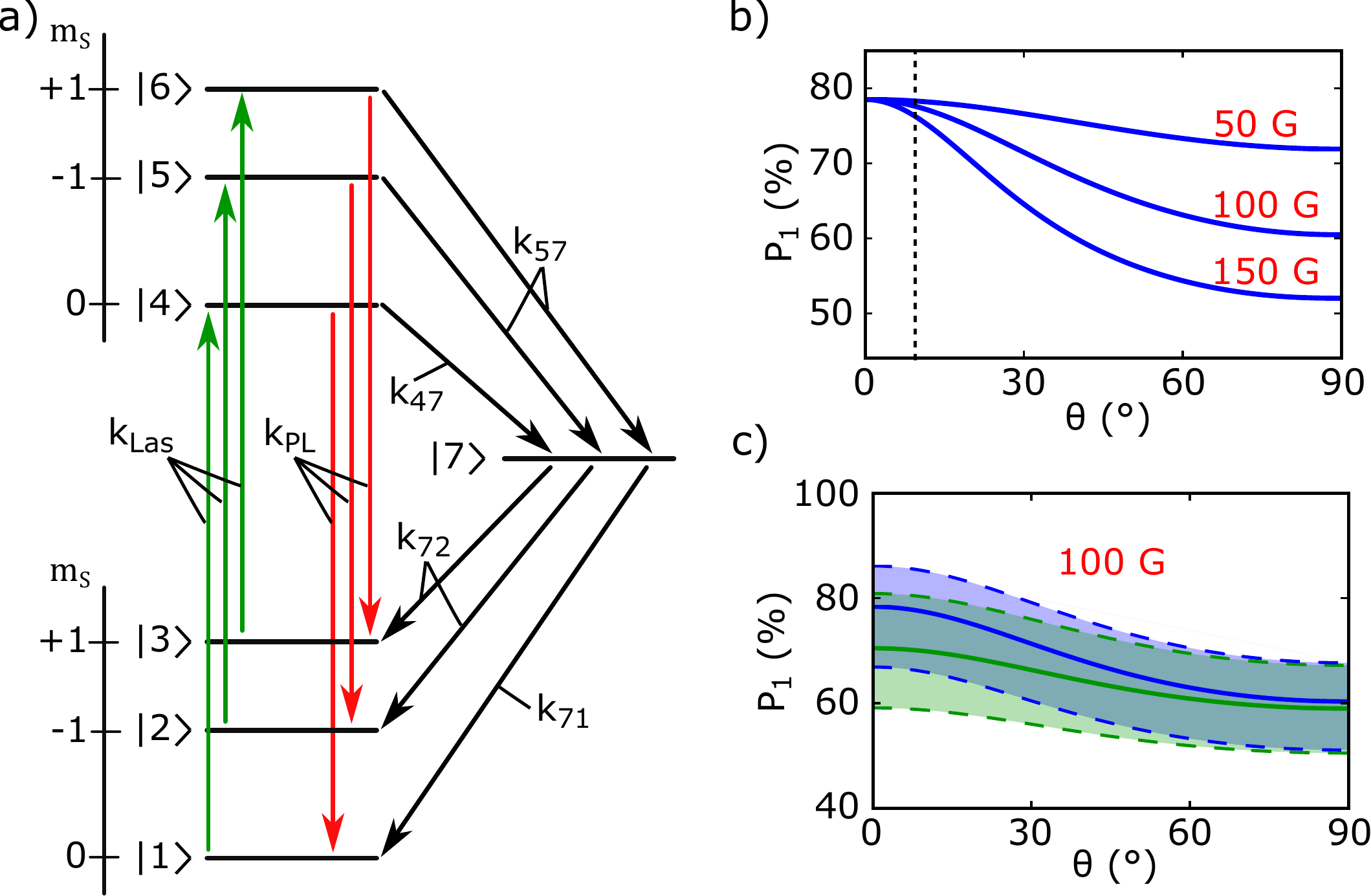}}
  \caption{a) NV center electron spin energy level structure. b) Population in the lowest energy state versus $\theta$ for B=50,100 and 150 G. The calculations are done with the rates given by the set 1 mean values (see Table 1). c) Population in the lowest energy state versus $\theta$ with B=100 G for the set 1 (in blue) and for the set 2 (in green). The dashed lines give the error margins (See Table 1).}
	\label{FigNVe}
\end{figure}

Before estimating the influence of the laser on the nuclear spin depolarisation rate in the presence of off-axis magnetic fielf, we describe the optical polarisation of the NV center electronic spin. We use the formalism of \cite{SITetienne_2012}. As shown in Fig.\ref{FigNVe}a), the relevant NV center energy level structure is composed of the spin-triplet ground state $^3A_2$ (states $\{\vert 1 \rangle,\vert 2 \rangle,\vert 3 \rangle\}$), the spin-triplet excited state $^3E$ (states $\{\vert 4 \rangle,\vert 5 \rangle,\vert 6 \rangle\}$) and an effective spin-singlet metastable state ($\vert 7 \rangle$). The spin-triplet excited state $^3E$ is an orbital doublet which is averaged at room temperature, its Hamiltonian $H_{e}^{es}/h = D^{es}\hat{S_z^2} + \gamma_e \hat{\textbf{S}}\cdot\textbf{B}$ is similar to the ground state one (with the same natural quantization axis) but with a different zero-field splitting value $D^{es} = 1.43$ GHz. The optical polarization of the electron spin in the $\vert m_S = 0 \rangle$ state relies on a spin-dependent relaxation of the optically exited states to the metastable state {\it via} a non-radiative inter-system crossing (ISC).

In a magnetic field that is along the NV center natural quantization axis, the natural basis set is $\{\vert i \rangle\}$. The incoherent transition rates between the seven levels resulting from optical excitation, photo-luminescence and non-radiative ISC, are conveniently defined in this basis. Optical excitation from the ground to the excited state and direct photo-luminescence from the excited to the ground state are assumed to be spin conservative and equal for the three electron spin states: $k_{14}=k_{25}=k_{36}=k_{Las}$ and $k_{41}=k_{52}=k_{63}=k_{PL}$. ISC transitions are characterized by $k_{47}\neq k_{57}=k_{67}$ and $k_{71}\neq k_{72}=k_{73}$. The five intrinsic parameters ($k_{PL}$,$k_{47}$,$k_{57}$,$k_{71}$,$k_{72}$) left in the problem have been determined experimentally. Their values are presented in table 1. Solving the corresponding rate equations shows that around 80$\%$ of the population is pumped into the state $\vert m_S = 0 \rangle$ of the spin triplet ground level (state $\vert 1 \rangle$).

\begin{table}
   \begin{tabular}{| m{2cm} || m{4cm} | m{4cm} | }
     \hline
     rate (MHz) & Set 1 & Set 2 \\ \hline
     $k_{PL}$ & 67.9$\pm$1.3 & 63$\pm$3\\ \hline
     $k_{47}$ & 5.7$\pm$0.7 & 12$\pm$3\\ \hline
		 $k_{57}$ & 49.9$\pm$1.6 & 80$\pm$6\\ \hline
		 $k_{71}$ & 1.01$\pm$0.28 & 3.3$\pm$0.4\\ \hline
		 $k_{72}$ & 0.75$\pm$0.11 & 2.4$\pm$0.4\\ \hline
   \end{tabular} \caption{Coupling rates between the NV center eigenstates. The values presented in the set 1 are taken from \cite{SITetienne_2012} where measurements were done on four different NV centers. The values quoted here corresponds to the weighted mean and standard error of the four measurements. The values of the set 2 are taken from \cite{SIPhysRevB.95.195308}, where measurements from \cite{SIRobledo_2011} were used.}
 \end{table}

In the presence of an off-axis magnetic field, states $\vert 1 \rangle$ to $\vert 7 \rangle$ are not eigenstates of the system because $\hat{H}_{e}$ and $\hat{H}_{e}^{es}$ are not diagonal in the NV center natural quantization basis. The new eigenstates $\{\vert \tilde{i} \rangle \}$ can be expressed as $\vert \tilde{i} \rangle = \sum_{j=1}^7{\alpha_{\tilde{i}j} \vert j \rangle}$ where the $\{\alpha_{\tilde{i}j}\}$ can be computed numerically by diagonalizing $\hat{H}_{e}$ and $\hat{H}_{e}^{es}$. A new set of rate equations for the states $\{\vert \tilde{i} \rangle \}$ can be obtained by calculating the new rates as $\tilde{k}_{\tilde{i}\tilde{j}} = \sum_{p=1}^7\sum_{q=1}^7\vert\alpha_{\tilde{i}p}\vert^2\vert\alpha_{\tilde{j}q}\vert^2k_{pq}$. In Fig.\ref{FigNVe}b), we show the population in the lowest energy state of the system as a function of the angle between the magnetic field and the NV center axis $\theta$ for  $B= 50, 100$ and 150 G. This shows that, for the magnetic field strength considered in this work, the NV electron spin optical polarization remains, for all $\theta$, on the same order of magnitude as for $\theta=0$ (longitudinal magnetic field). On this graph the orientation for which CPT and nuclear spin polarization experiments have been realized is indicated by the vertical dashed line.

In Fig.\ref{FigNVe}c), we plot the same curve as above for $B=100$ G using the two available set of parameters, with their experimental uncertainties. It is noteworthy that, using currently available experimental measurements, the value for the optical polarization achievable for NV centers at room temperature is determined with a level of uncertainty of around 10$\%$.

\begin{figure}[!ht] 
  \centering \scalebox{0.6}{\includegraphics{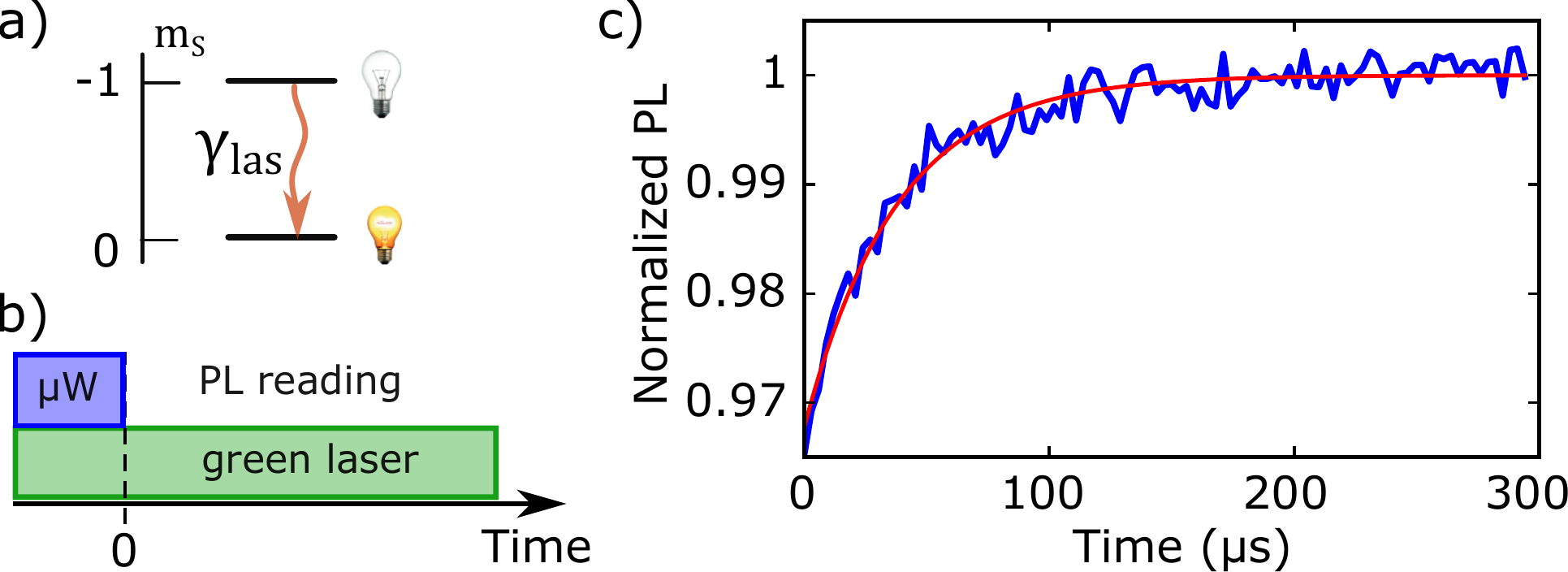}}
  \caption{a) Phenomenological picture for the NV electron spin polarization. b) Experimental sequence used to determine $\gamma_{Las}$. c) Typical measurements of $\gamma_{Las}$, data are in blue, solid red line is an exponential fit to the data.}
	\label{gammaLas}
\end{figure}

A very simplified picture can be used to describe the effect of optical illumination on the NV electron spin ground level by modeling the polarisation via a relaxation from the states $\vert m_S \pm 1 \rangle$ towards the state $\vert m_S = 0 \rangle$ at a rate $\gamma_{Las}$ (see Fig.\ref{gammaLas}a)). Experimentally, $\gamma_{Las}$ is obtained by monitoring the PL counts rate temporal evolution after switching off a microwave field resonant with an electron transition (see Fig.\ref{gammaLas}b)). While the microwave field populates the $\vert m_S = -1 \rangle$ (or $\vert m_S = 1 \rangle$) state, population returns to the $\vert m_S = 0 \rangle$ after the field is switched off. As shown in Fig.\ref{gammaLas}c), fitting the temporal evolution of the PL by an exponential curve gives $\gamma_{Las}$.

\subsubsection{Electron-nuclear spin state mixing under off-axis magnetic field}

In this section, we give describe in more details the electron-nuclear spin coupling induced by the presence of an off-axis magnetic field.
We first recall the NV-center Hamiltonian as given in the main text
$$ \hat{H}_{NV}/h = D\hat{S_z^2} + \gamma_e \hat{\textbf{S}}\cdot\textbf{B} + Q\hat{I_z^2} - \gamma_n \hat{\textbf{I}}\cdot\textbf{B} + \underline{\underline{\mathcal{A}}} \hat{\textbf{I}}\cdot\hat{\textbf{S}}.$$

\noindent
This Hamiltonian is expressed in the NV center natural quantization basis (the $z$-axis is the NV center axis). Without loss of generality, we take the magnetic field (of magnitude $B$) in the $z-x$-plane, forming an angle $\theta$ with the $z$-axis and write
$$ \gamma_e \hat{\textbf{S}}\cdot\textbf{B} = \gamma_e B_{\vert\vert} \hat{S_z}+\gamma_e B_{\perp}\hat{S_x},$$
$$ \gamma_n \hat{\textbf{I}}\cdot\textbf{B} = \gamma_n B_{\vert\vert}\hat{I_z}+\gamma_n B_{\perp}\hat{I_x},$$

\noindent
with $B \cos \theta = B_{\vert\vert}$ and $B \sin \theta = B_{\perp}$.

\begin{figure}[!ht] 
  \centering \scalebox{0.7}{\includegraphics{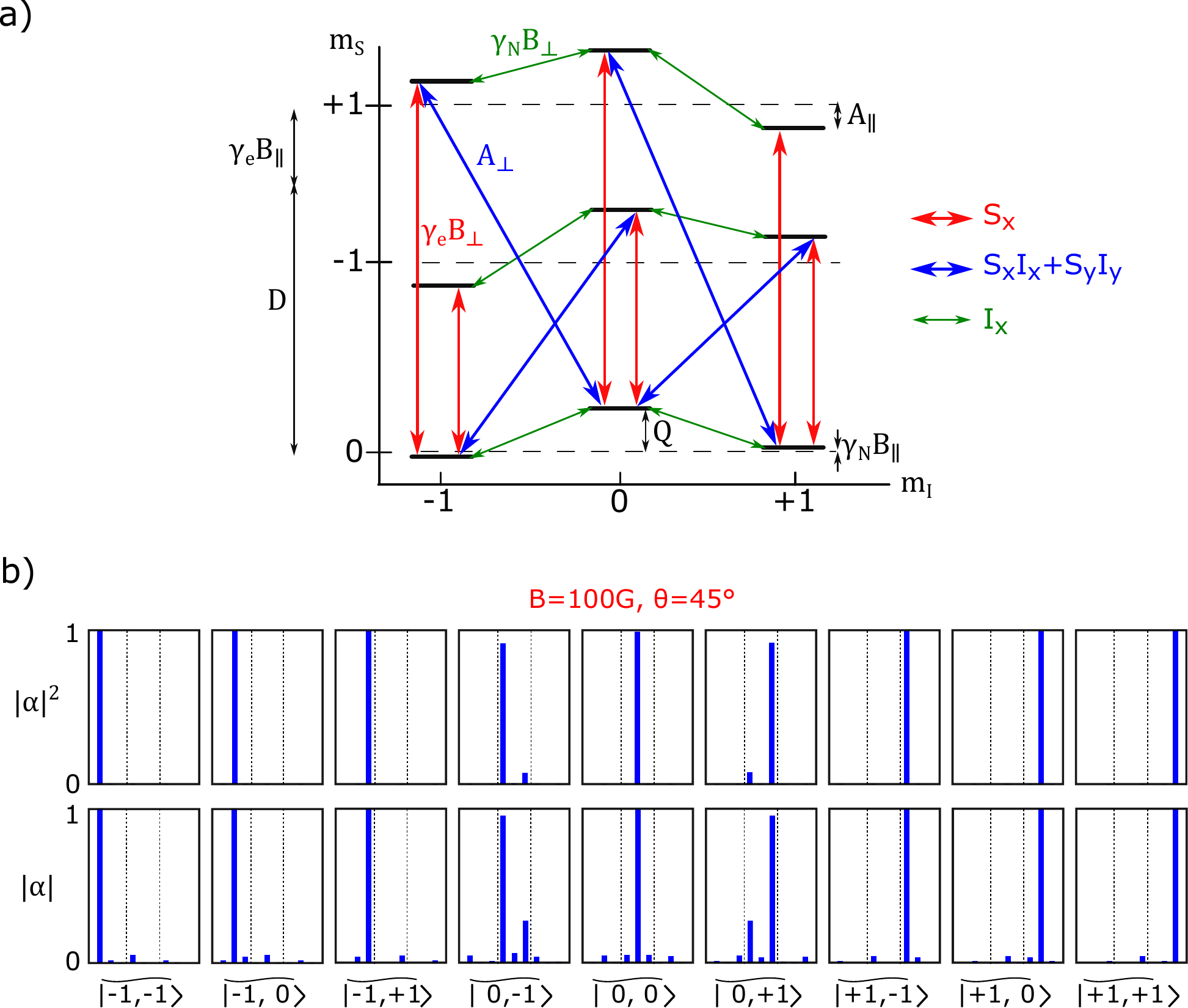}}
  \caption{a) Energy levels in the NV center natural basis, the arrows represent the mixing due to the Hamiltonian off-diagonal terms. b) Decomposition of the Hamiltonian eigenstates (denoted as $\widetilde{\vert m_S,m_I \rangle}$) onto the NV center natural basis, for $B$=100 G and $\theta$=45$^{\circ}$. Both the amplitude (bottom) and amplitude squared (top) of the $\alpha$ coefficients are represented. For each eigenstate $\widetilde{\vert m_S,m_I \rangle}$, the blue bars represent the decomposition onto the 9 states $\vert m_S,m_I \rangle$ with the same ordering as the new eigenstates.}
	\label{Fig-en-1}
\end{figure}

In Fig.\ref{Fig-en-1}a), we show a graphical representation of this Hamiltonian, highlighting the off-diagonal terms responsible for the electron-nuclear spin state mixing.
We can formally express the eigenstates of $\hat{H}_{NV}$ as $\vert i \rangle = \sum_{m_S,m_I} \alpha_{m_S,m_I}^{i}\vert m_s,m_I \rangle$, where $\{\vert m_s,m_I \rangle\}_{m_S=-1,0,+1,m_I=-1,0,+1}$ denote the quantum states in the NV center basis.
We plot in Fig.\ref{Fig-en-1}b) $\vert \alpha_{m_S,m_I}^{i} \vert$ and $\vert \alpha_{m_S,m_I}^{i} \vert^2$ for a magnetic field $B$=100G and $\theta$=45$^{\circ}$. As the state mixing remains small, each eigenstate $\vert i \rangle$ is mainly composed of 1 bare state $\vert m_s,m_I \rangle$ that we label $\widetilde{\vert m_S,m_I \rangle}$.

The only pair of states that are significantly mixed by $\hat{H}_{NV}$ are $\vert 0,-1 \rangle$ and $\vert 0,+1 \rangle$ in the ground state of the electronic spin because they have very little energy difference ($2\gamma_nB_{\vert\vert}$). The small mixing between, for example, the states $\vert 0,0 \rangle$ and $\vert 0,+1 \rangle$ is responsible for the nuclear spin exchanging transition in the electron spin resonance spectrum. Interestingly, given that $\gamma_n \ll \gamma_e, D/B$, the contribution of the $I_x$ term in the Hamiltonian (green arrows on the Fig.\ref{Fig-en-1}a)) is negligible. The electron-nuclear spin state mixing at stake in this work originates mostly from both the hyperfine interaction term $\mathcal{A}_{\perp}(\hat{S_x}\hat{I_x}+\hat{S_y}\hat{I_y})$ and the magnetic term $\gamma_e B_{\perp}\hat{S_x}$ that involves the transverse component of the magnetic field. As can be seen in Fig.\ref{Fig-en-1}a), mixing between the relevant states are a second order perturbation with a $\Lambda$-scheme involving one $\hat{S_x}$ and one $(\hat{S_x}\hat{I_x}+\hat{S_y}\hat{I_y})$ coupling. 
For example the states $\vert 0,0 \rangle$ and $\vert 0,+1 \rangle$ are mixed through two of these $\Lambda$-schemes via the $\vert -1,+1 \rangle$ and $\vert +1,0 \rangle$ states.

\begin{figure}[!ht] 
  \centering \scalebox{0.64}{\includegraphics{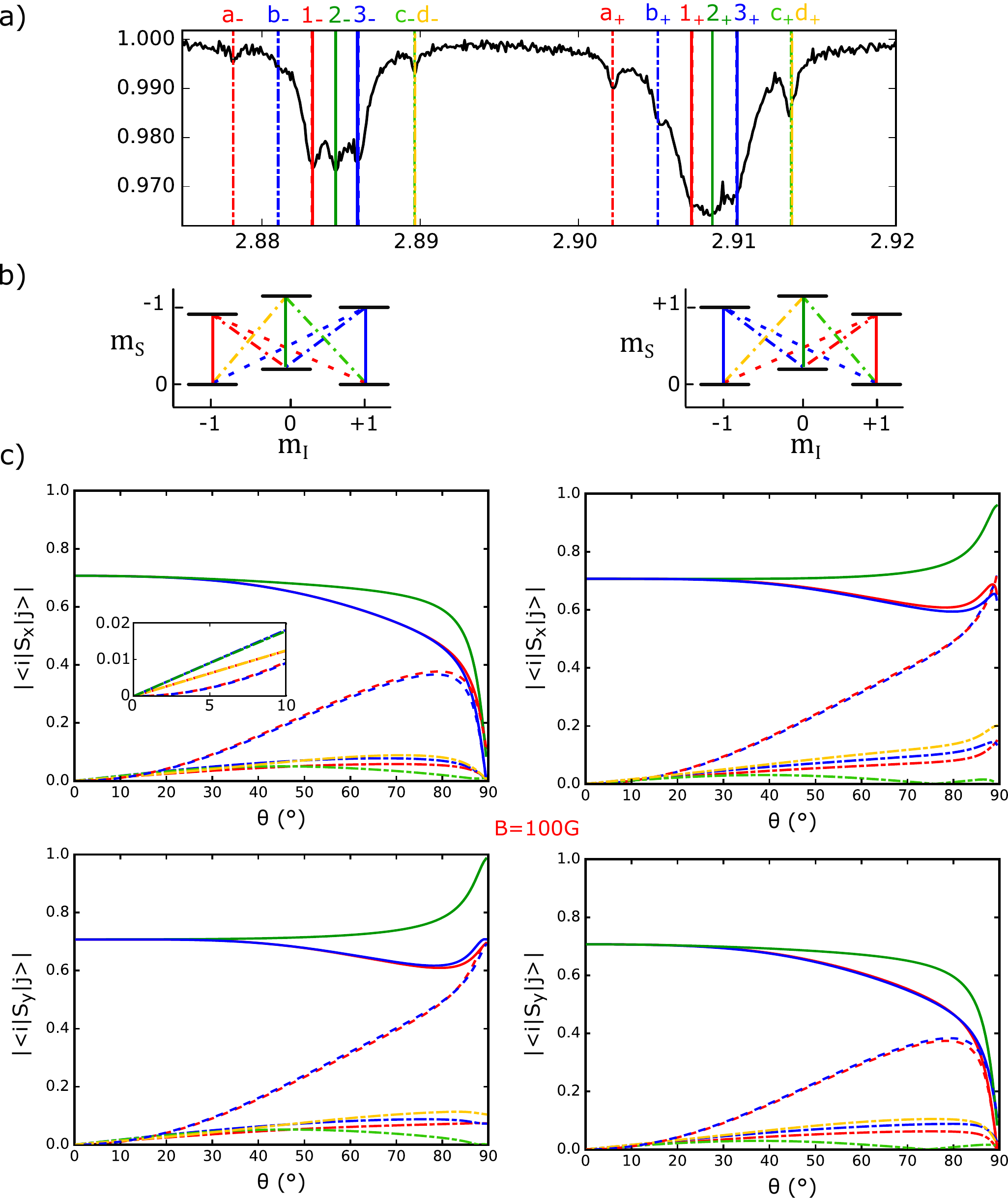}}
  \caption{a) ODMR spectrum taken in the presence of a magnetic field of $B$=80.8 G and $\theta$=88.0$^{o}$. Black solid line is the data, colored vertical lines are the calculated transition frequencies. b) Energy level diagrams for the $m_S = 0,-1$ (left) and $m_S=0,+1$ (right) states. The allowed transitions between states are indicated by the colored lines with the same color code as in a). c) Calculated transition strengths versus $\theta$ with $B$=100 G. The $\vert m_S = 0 \rangle \leftrightarrow \vert m_S = -1(+1) \rangle$ transitions are on the left (right), panels on the top (bottom) are with a microwave field polarized along the $x$-axis ($y$-axis). The same color code as in b) is used. The inset on the top left panel is a zoom on the low angle values.}
	\label{Fig-en-2}
\end{figure}

\begin{figure}[!ht] 
  \centering \scalebox{0.7}{\includegraphics{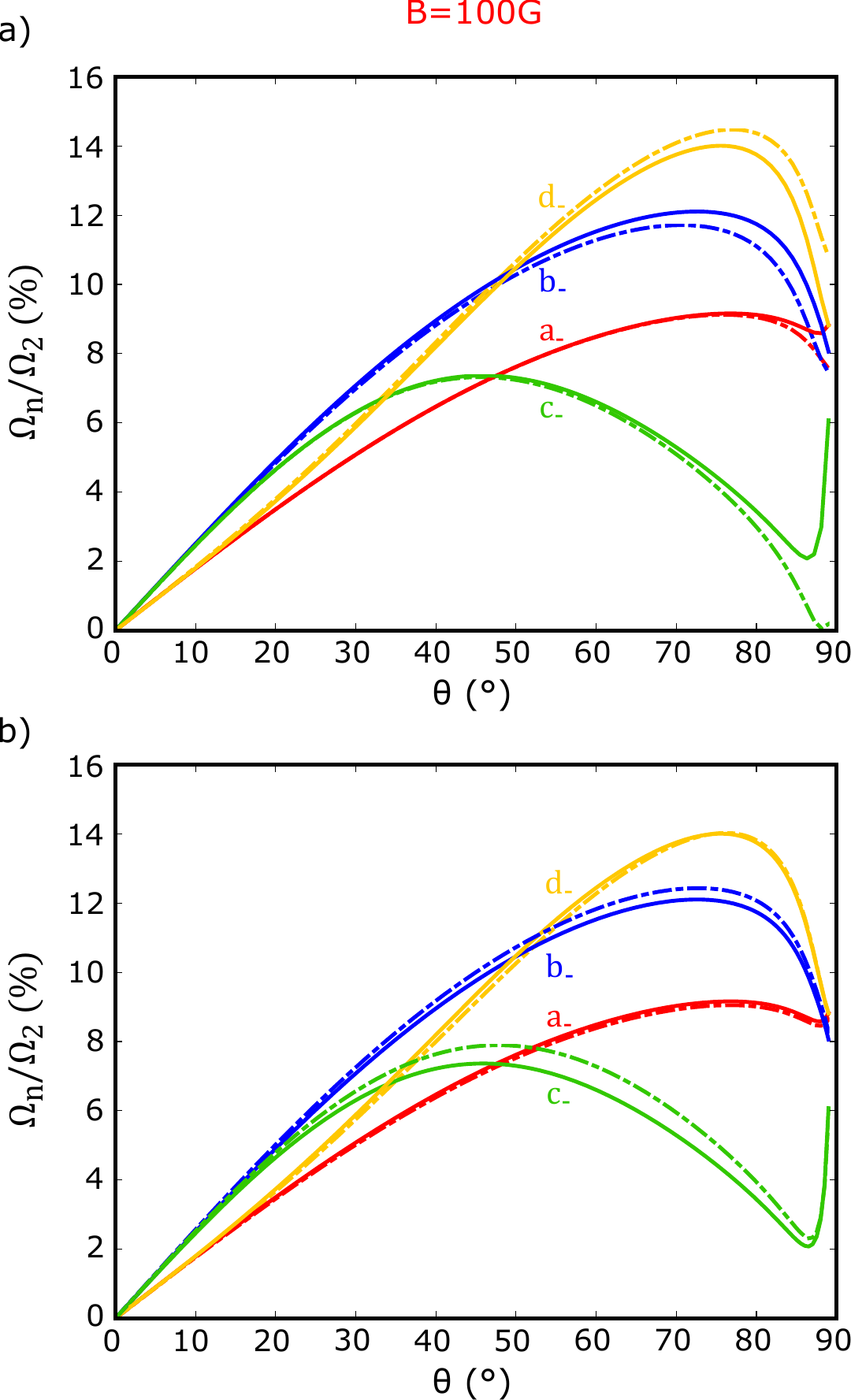}}
  \caption{a) Strength of the $a_-$, $b_-$, $c_-$ and $d_-$ transitions over the one of the transition $2_-$ for a driving microwave field polarized along both the $x$- (solid lines) and the $y$-axis (doted-dashed lines). b) Strength of the $a_-$, $b_-$, $c_-$ and $d_-$ transitions over the one of the transition $2_-$ for a driving microwave field polarized along the $x$-axis, with (solid lines) and without (doted-dashed lines) the term $\gamma_n B_{\perp}\hat{I_x}$ in the Hamiltonian.}
	\label{Fig-en-3}
\end{figure}

We now analyse the strength of the different transitions that can be driven with a magnetic field of magnitude $B_{\mu W}$ oscillating in the microwave regime. These transitions are shown in Fig.\ref{Fig-en-2}a) and b) which shows experimental data and the corresponding levels in the eigenbasis of $S_z$.

In Fig.\ref{Fig-en-2}c), we plot the quantity $\vert \langle i \vert S_{x(y)} \vert j \rangle \vert$ as a function of $\theta$. It corresponds to the $\vert i \rangle \leftrightarrow \vert j \rangle$ transition strength normalized by $\gamma_e B_{\mu W}$, for a microwave field polarized along the $x$-axis ($y$-axis).
This plot was taken with a magnetic field of $B$=100 G.
The weakly allowed transition where the nuclear spin is changed by one quantum (labelled $a$, $b$, $c$ and $d$ in the main text) increases linearly with $\theta$, like the $\gamma_e B_{\perp}\hat{S_x}$ coupling term, and reaches values on the order of 10$\%$ of the nuclear spin preserving transition.
Interestingly, the transitions where the nuclear spin is changed by two quanta have, for a $\theta$ close to $\pi/2$, a strength comparable to the ones of nuclear spin preserving transition. This is due to the strong mixing between the $\vert 0,-1 \rangle$ and $\vert 0,+1 \rangle$ states. At small $\theta$, their strengths vary quadratically with $\theta$, because the states $\vert 0,-1 \rangle$ and $\vert 0,+1 \rangle$ are mixed through a fourth order mixing involving two $\gamma_e B_{\perp}\hat{S_x}$ coupling terms. It noteworthy that these transitions occur at roughly the same frequency than the spin preserving transitions with a frequency difference of $\gamma_nB_{\vert\vert}$, \textit{i.e.} on the order of 10 kHz for the magnetic field considered in this work, and cannot be distinguished from the latter.
Finally, we note that, when $\theta$ approaches 90$^{\circ}$, either the $\vert m_S = 0 \rangle \leftrightarrow \vert m_S = -1 \rangle$ or the $\vert m_S = 0 \rangle \leftrightarrow \vert m_S = +1 \rangle$ transition vanishes, at the benefit of the other, when the driving microwave field is polarized along the $x$- or the $y$-axis.

To compare experimental measurements with these theoretical predictions, it is convenient to calculate the transition strength ratio between the nuclear-spin exchanging transitions and the nuclear spin preserving transition. In Fig.\ref{Fig-en-3}a), we plot the strength of the $a_-$, $b_-$, $c_-$ and $d_-$ transitions divided by the strength of transition $2_-$, for a driving microwave field polarized along both the $x$- and the $y$-axis. While the two polarization leads to little differences for all transitions, we note that the smallest difference is for the $a_-$ transition. This is the one that has been compared with experimental data in the main text.

Finally, we plot in Fig.\ref{Fig-en-3}b) the strength of the $a_-$, $b_-$, $c_-$ and $d_-$ transitions over the one of the transition $2_-$ for a driving microwave field polarized along the $x$-axis, with and without the term $\gamma_n B_{\perp}\hat{I_x}$ in the Hamiltonian. This shows that, as mentioned above, this term is almost negligible for the electron-nuclear spin state mixing investigated in this work.

\subsubsection{NV center electron-nuclear spins photo-physics}

In this section we discuss theoretical modeling of the dynamics of the NV center coupled electron-nuclear spin system in the presence of optical illumination and off-axis magnetic field \cite{SIPhysRevB.92.020101,SIPhysRevB.89.205202,SIPhysRevB.95.195308,SIPhysRevLett.124.153203}. Numerical resolution of this dynamics can be compared with the experimental measurements of the nuclear spin depolarization rates versus electron spin polarization rate presented in fig.3e in the main text.

As discuss in a previous section (see fig.\ref{FigNVe}a), the NV center electron spin photo-physics can be described with a 7-level system (3 ground levels, 3 excited levels and 1 metastable level) connected through incoherent transition rates and with Hamiltonian dynamics within the 3 ground states ($\hat{H}_e$) and the 3 excited states ($\hat{H}_e^{es}$). The $I=1$ $^{14}$N nuclear spin will now be included in the electronic level scheme. The Hamiltonian of excited electronic spin state is then
$$ \hat{H}_{NV}^{es}/h = D^{es}\hat{S_z^2} + \gamma_e \hat{\textbf{S}}\cdot\textbf{B} + Q\hat{I_z^2} - \gamma_n \hat{\textbf{I}}\cdot\textbf{B} + \underline{\underline{\mathcal{A}}}^{es} \hat{\textbf{I}}\cdot\hat{\textbf{S}},$$

\noindent
where $\underline{\underline{\mathcal{A}}}^{es}$ is the excited state diagonal hyperfine interaction tensor with $\mathcal{A}_{zz}^{es} = 40$ MHz \cite{SIPhysRevB.81.035205} and $\mathcal{A}_{xx}^{es} = \mathcal{A}_{yy}^{es} = \mathcal{A}_{\perp}^{es} = 23(3)$ MHz \cite{SIPhysRevB.95.195308}. The incoherent transitions, that have been defined in the NV natural quantization basis for the electron spin, can be straightforwardly extended to the electron-nuclear spin system in the same basis by assuming that they preserve the nuclear spin. Each of the 12 incoherent transitions (see fig.\ref{FigNVe}a) is split into 3 transitions with the same rate.

The time evolution of the resulting density matrix $\rho$ is given by
$$ \frac{d}{dt}\rho = -\frac{i}{\hbar}\left[\hat{H},\rho\right] + \hat{L}\left[\rho\right],$$
where
$$ \hat{L}\left[\rho\right] = \sum_{k=1}^{36}\left( L_k\rho L_k^{\dagger} - \frac{1}{2} L_k^{\dagger}L_k\rho - \frac{1}{2} \rho L_k^{\dagger}L_k \right), $$

\noindent
The Lindbald operator $\hat{L}$ describes the 36 incoherent transitions through their jump operator, $L_k = \sqrt{\Gamma_{nm}}\vert m \rangle\langle n \vert$ describing a transition from state $\vert n \rangle$ to state $\vert m \rangle$.

To compare this model with the experimental measurements of the nuclear spin depolarization rates shown in fig.3e in the main text we solve this dynamics numerically. 
We start in the $\widetilde{\vert 0,0 \rangle}$ and the $\widetilde{\vert 0,+1 \rangle}$ states (eigenstates of $\hat{H}_{NV}$ which are mainly composed of the $\vert m_S=0,m_I=0 \rangle$ and $\vert m_S=0,m_I=+1 \rangle$ states, respectively, as defined in the previous section). 
We can extract the total population of the 9 excited levels $P_{exc}$, which is proportional to the NV center PL and also the population difference $P_{Signal}$ between the $\widetilde{\vert 0,0 \rangle}$ and the $\widetilde{\vert -1,0 \rangle}$ states (and between the $\widetilde{\vert 0,+1 \rangle}$ and the $\widetilde{\vert -1,+1 \rangle}$ states). The latter is proportional to the signal that we measure experimentally in fig.3e of the main text, when applying a resonant microwave field.

\begin{figure}[!ht] 
  \centering \scalebox{0.7}{\includegraphics{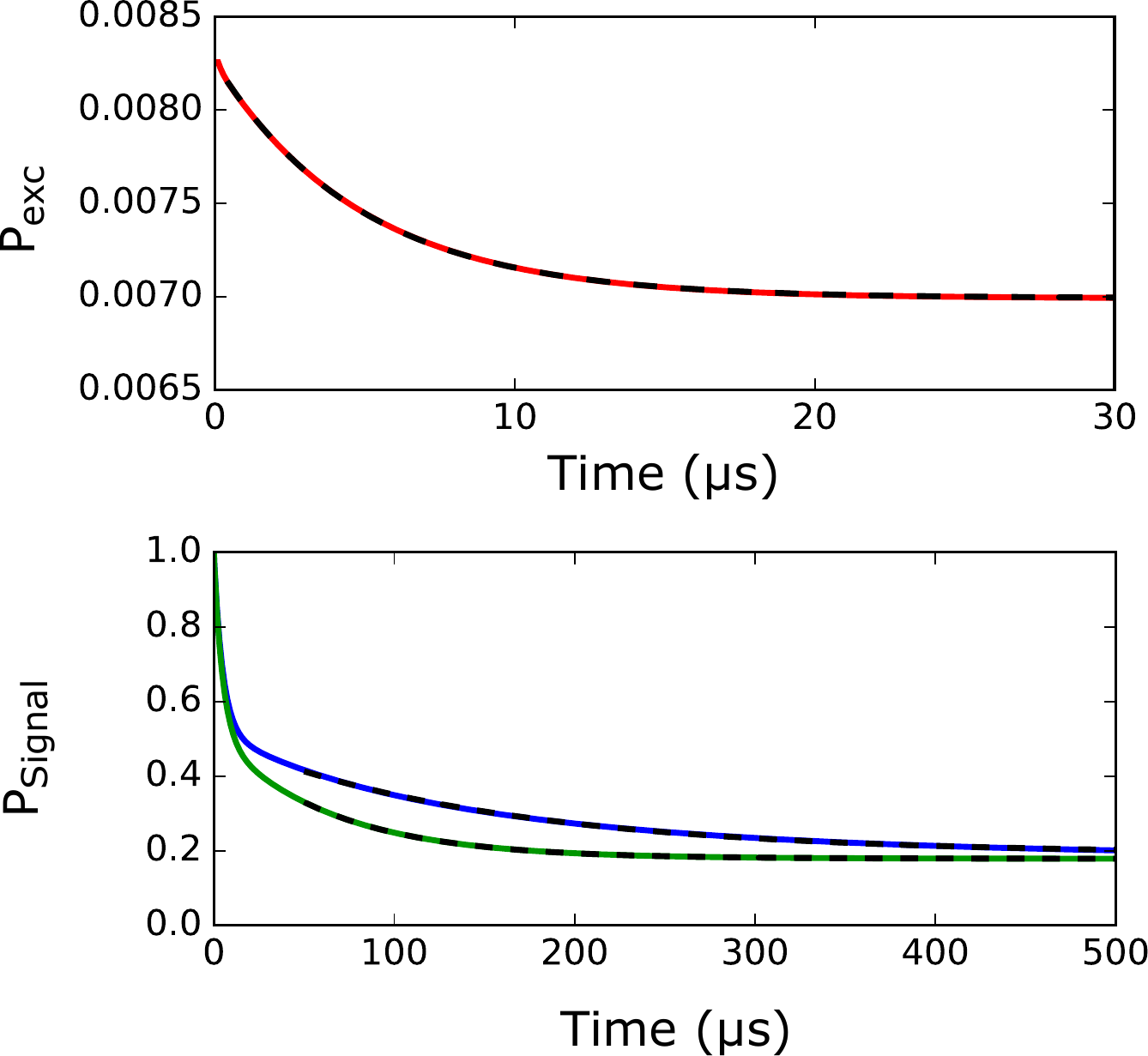}}
  \caption{Top: total population in the 9 excited levels $P_{exc}$ versus time for $\vert \Psi(t=0) \rangle = \widetilde{\vert 0,0 \rangle}$. Bottom, population difference between the $\widetilde{\vert 0,0 \rangle}$ and the $\widetilde{\vert -1,0 \rangle}$ states for $\vert \Psi(t=0) \rangle = \widetilde{\vert 0,0 \rangle}$ (green) and between the $\widetilde{\vert 0,+1 \rangle}$ and the $\widetilde{\vert -1,+1 \rangle}$ states for $\vert \Psi(t=0) \rangle = \widetilde{\vert 0,+1 \rangle}$ (blue). This was calculating with $(B,\theta)=(82.71~{\rm G}, 10.2^{\circ}$) as in the experimental curve of Fig.3e, and using $k_{Las} = 0.01\times k_{PL}$. The dashed black curves are exponential decay fits.}
	\label{depol_fit}
\end{figure}

In fig.\ref{depol_fit}, we plot $P_{exc}$ and $P_{Signal}$ as a function of time for different initial states, using the transition rate values from \cite{SITetienne_2012} (set 1 in table 1). Fitting these curves with exponential decays allow to get values for the electron spin polarization rate ($\gamma_{Las}$) and the nuclear spin depolarization rate $\gamma_{las}^{\widetilde{\vert m_I= +1\rangle}}$ and $\gamma_{las}^{\widetilde{\vert m_I= 0\rangle}}$,  that can be compared to experimental data.

\begin{figure}[!ht] 
  \centering \scalebox{0.7}{\includegraphics{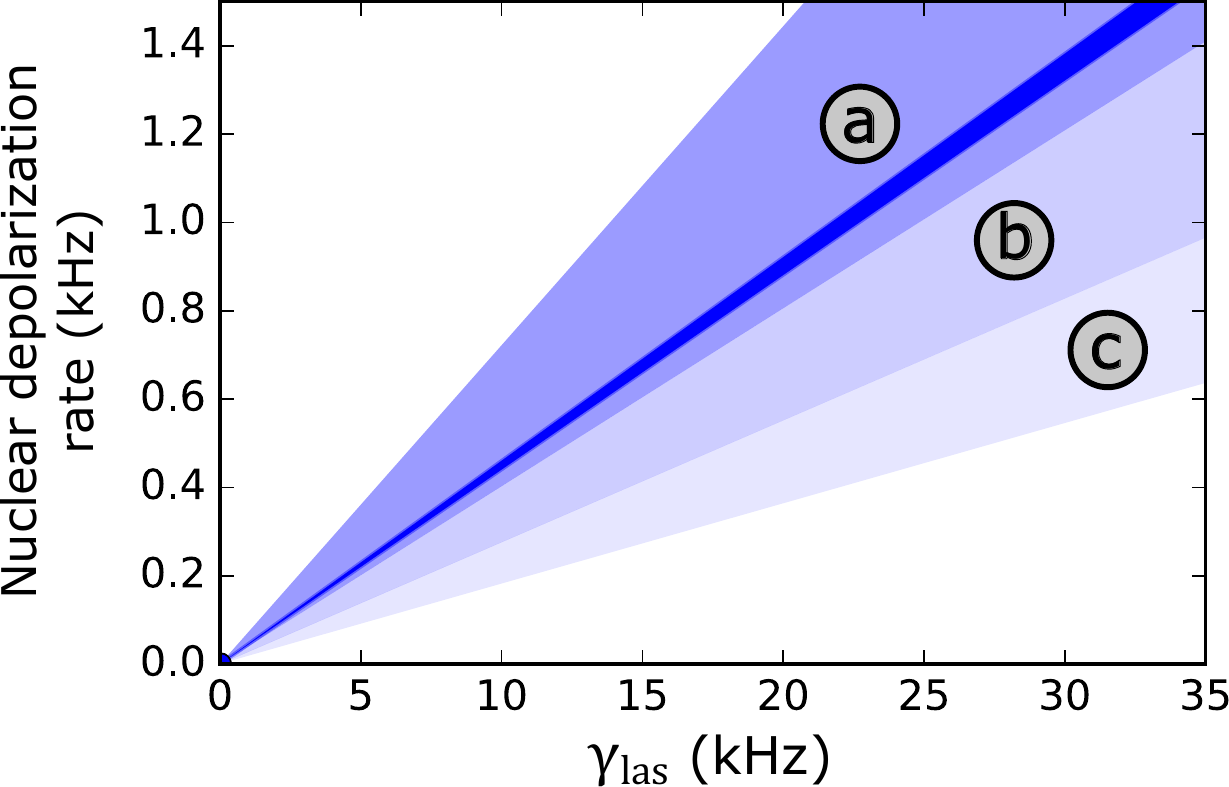}}
  \caption{Nuclear spin depolarization rate versus $\gamma_{Las}$ including parameters uncertainty. Region $a$ corresponds to uncertainty on the hyperfine interaction tensor transverse component, region $b$ corresponds to the difference between the two available experimental sets for the transitions strengths, region $c$ to the uncertainty within one experimental set}
	\label{nucdepol}
\end{figure}

In fig.\ref{nucdepol} we plot the nuclear spin depolarization rate of the state $\widetilde{\vert 0,+1 \rangle}$ versus $\gamma_{Las}$ for the three identified sources of uncertainty in the physical parameters, {\it i.e}, the value of the hyperfine interaction tensor transverse component $\mathcal{A}_{\perp}^{es} = 23(3)$ MHz, the available experimental sets for the transitions strengths parameters and their own uncertainty. We note that overall, $\gamma_{las}^{\widetilde{\vert m_I= +1\rangle}}$ and $\gamma_{las}^{\widetilde{\vert m_I= 0\rangle}}$ are currently only loosely bounded by the available experimental parameters. However, their ratio $\gamma_{las}^{\widetilde{\vert m_I= +1\rangle}} / \gamma_{las}^{\widetilde{\vert m_I= 0\rangle}}$ is almost constant for most of the parameters. The discrepancy between the theoretical and experimental values, 2.27(3) and 1.8(1), respectively, remains to be explained. It could be due to the presence of electron spin non-preserving radiative transitions \cite{SIRobledo_2011} or photo-induced excitation to the neutral charge NV state, which are not accounted for in the present model.

\end{document}